\documentclass[fleqn,usenatbib]{mnras}

\usepackage{newtxtext,newtxmath}

\usepackage[T1]{fontenc}

\usepackage[table,x11names]{xcolor}
\usepackage{amsfonts,amsmath,graphicx,natbib,subfigure,color,gensymb,longtable,footnote}
\usepackage{subfigure,color,gensymb,longtable,color,xcolor,soul}

\DeclareRobustCommand{\VAN}[3]{#2}
\let\VANthebibliography\thebibliography
\def\thebibliography{\DeclareRobustCommand{\VAN}[3]{##3}\VANthebibliography}

\usepackage{graphicx}	% Including figure files
\usepackage{amsmath}	% Advanced maths commands
\title[X-ray and Radio Observations from SN\,2014dt]{Constraints on the Sub-pc Environment of the Nearby Type Iax SN\,2014dt from Deep X-ray and Radio Observations}

\author[C.~M.~Stauffer et al.]{
C.~M.~Stauffer,$^{1}$\thanks{E-mail: cstauffer@u.northwestern.edu}
R.~Margutti,$^{1}$
J.~D.~Linford,$^{2}$
L.~Chomiuk,$^{3}$
D.~L.~Coppejans,$^{1}$
L.~Demarchi,$^{1}$ \newauthor
W.~Jacobson-Gal\'an,$^{1}$
J.~Bright,$^{1}$
R.~J.~Foley,$^{4}$
A.~Horesh$^{5}$
A.~Baldeschi$^{1}$ 
\\
$^{1}$ Center for Interdisciplinary Exploration and Research in Astrophysics (CIERA) and Department of Physics and Astronomy,\\ 
Northwestern University, Evanston, IL 60208, USA\\
$^{2}$National Radio Astronomy Observatory New Mexico Array Operations Center,1003 Lopezville Rd, Socorro, NM 87801, USA \\
$^{3}$Department of Physics and Astronomy, Michigan State University, 567 Wilson Road, East Lansing, MI 48824, USA\\
$^{4}$Department of Astronomy and Astrophysics, University of California, Santa Cruz, CA 95064, USA\\
$^{5}$Racah Institute of Physics, The Hebrew University of Jerusalem, Jerusalem 91904, Israel }

\date{Accepted XXX. Received YYY; in original form ZZZ}

\pubyear{2015}

\begin{document}
\label{firstpage}
\pagerange{\pageref{firstpage}--\pageref{lastpage}}
\maketitle

\begin{abstract}
We present X-ray and radio observations of what may be the closest type Iax supernova (SN) to date, SN\,2014dt ($d=12.3-19.3$ Mpc), and provide tight constraints on the radio and X-ray emission. We infer a specific radio luminosity $L_R<(1.0-2.4)\times 10^{25}\,\rm{erg\,s^{-1}\,Hz^{-1}}$ at a frequency of 7.5 GHz and a X-ray luminosity $L_X<1.4\times 10^{38}\,\rm{erg\,s^{-1}}$ (0.3-10 keV) at $\sim38-48$ days post-explosion.  We interpret these limits in the context of Inverse Compton (IC) emission and synchrotron emission from a population of electrons accelerated at the forward shock of the explosion in a power-law distribution $N_e(\gamma_e)\propto \gamma_e^{-p}$ with $p=3$.  Our analysis constrains the 
progenitor system mass-loss rate to be $\dot M<  5.0 \times10^{-6} \rm{M_{\sun}\,yr^{-1}} $ at distances $r\lesssim 10^{16}\,\rm{cm}$ 
for an assumed wind velocity  $v_w=100\,\rm{km\,s^{-1}}$, and a fraction of post-shock energy into magnetic fields and relativistic electrons of $\epsilon_B=0.01$ and $\epsilon_e=0.1$, respectively. 
This result rules out some of the parameter space of symbiotic giant star companions, and it is consistent with the low mass-loss rates expected from He-star companions. 
Our calculations also show that the improved sensitivity of the next generation Very Large Array (ngVLA) is needed to probe the very low-density media characteristic of He stars that are the leading model for binary stellar companions of white dwarfs giving origin to type Iax SNe.
\end{abstract}


%%%%%%%%%%%%%%%%%%%%%%%%%%%%%%%%%%%%%%%%%%%%%%%%%%

%%%%%%%%%%%%%%%%% BODY OF PAPER %%%%%%%%%%%%%%%%%%

\section{Introduction}
\label{S:introduction}

Type Iax supernovae (SNe) are a peculiar class of thermonuclear explosions. After the first member was identified in 2002 \citep{li2003sn}, dozens of members of the type Iax supernova class have since been discovered \citep{Foley13}. Observationally, type Iax SNe show clear spectroscopic similarities with type Ia SNe at maximum light (\citealt{li2003sn,branch2005comparative,chornock2006spectropolarimetry,foley2009sn,foley2009early,phillips2007peculiar}), which suggests a physical connection to thermonuclear explosions of C/O (or C/O/Ne) white dwarfs (WDs) in binary systems (see \citealt{Jha2017} for a recent review). However, a key physical difference with respect to type Ia SNe is that SNe Iax may not always completely destroy the WD progenitor star, but may instead leave behind a bound remnant (e.g., \citealt{Foley_2010,Jordan12,Kromer13, Fink14, Long14}), as was proposed for SN\,2012Z (\citealt{McCully2014,foley2009early,Foley2016}).
A fundamental question in the study of SNe Iax is the nature of the companion star of the primary WD. In this paper, we present deep X-ray and radio observations of one of the closest type Iax SN to date, SN\,2014dt ($d\sim$ 12.3-19.3 Mpc; \citealt{Foley2015a, Fox2015}). These observations provide tight constraints on the density of the nearby environment of the explosion, which was sculpted by the recent mass-loss history of the companion star, enabling insight into the nature of the progenitor system.

Unlike type Ia SNe, which show maximum light ejecta expansion velocities of $v\sim$10000 $\rm{km} \ \rm{s^{-1}}$, type Iax SNe typically have lower ejecta velocities (2000 $\rm{km\,s^{-1}}$ $\lesssim v \lesssim$ 8000 $\rm{km\,s^{-1}}$; e.g., \citealt{foley2009sn,McCully2014,Foley13,mccully2014hubble,narayan2011displaying,white2015slow}), which, on average, suggests a less energetic explosion than type Ia SNe (however \citealt{Stritzinger15} inferred an explosion kinetic energy $E_k\sim 10^{51}$ erg for SN\,2012Z, similar to SNe Ia). SNe Iax also show lower peak  optical luminosities corresponding to absolute V-band magnitudes in the range -13 $\gtrsim M_{V}$ $\gtrsim$ -19 mag, and typically have faster rise times to maximum light when compared to type Ia SNe (\citealt{Magee16,Magee17,Jha2017,mcclelland2010,stritzinger2014,stritzinger2014optical}), which points at lower ejecta  mass and $^{56}\rm{Ni}$ mass. Additionally, the optical spectra of SNe Iax never become dominated by broad forbidden lines at late times (i.e., never become fully ``nebular'', e.g.,  \citealt{Sahu08,McCully2014,Yamanaka15,Foley2016}). Finally, SNe Iax show a marked preference for late-type host galaxies, suggesting a connection to a relatively young stellar population (\citealt{Foley_2010,Lyman2013,Lyman2018,Li2017,takaro2020}). Currently, only one SN Iax (SN\,2008ge) has been discovered in an S0-type galaxy, with no evidence for star formation or massive stars \citep{Foley_2010}. 

The nature of the stellar companions to WDs that produce SNe Iax is largely unknown.  The leading paradigm is that these companions are He-stars, as originally postulated by \citet{Foley13}, and supported by binary simulations (e.g.,  \citealt{wang2013}). From an observational perspective, the notion of He-star companions is supported by the first detection of a progenitor system of a WD explosion for the type Iax SN\,2012Z, reported by  \citet{McCully2014}. These authors find a luminous blue source in pre-explosion Hubble Space Telescope (HST) images at the location of SN\,2012Z, which is consistent with the presence of a He-star companion.\footnote{He-star companion is the favored explanation but an accretion disk around a WD is also a viable interpretation. While the HST data alone are consistent with a massive star as well, the combination with other data makes this scenario no longer viable \citep{McCully2014}.} 
Currently, SNe 2008ge, 2012Z, and 2014dt are the only 
SNe Iax with  deep pre-explosion  imaging. No progenitor system was detected for either SN\,2008ge \citep{Foley_2010} or SN\,2014dt  \citep{Foley2015a}. For SN\,2008ge HST images ruled out the most massive stars as progenitors with initial masses $M < 85 M_{\sun}$ for main sequence stars and $M < 12 M_{\sun}$ for horizontal and red giant branch stars \citep{foley2010progenitor}. 

In the specific case of SN\,2014dt, the classes of stellar companions ruled out by HST pre-explosion imaging include the following \citep{Foley2015a}: (i) main sequence stars with initial mass $M > 16\,\rm{M_{\odot}}$ and (ii) red giant/horizontal branch (RG/HB)  stars with $M> 8\,\rm{M_{\odot}} $. \citet{Foley2015a} also conclude that Wolf-Rayet (WR) stars are unlikely companions due to the absence of other nearby bright stars at the location of SN\,2014dt. To complement this work, we study the nearby environment of SN\,2014dt which provides an indirect probe of its progenitor system.

SN\,2014dt was discovered on 2014 October 29.84 UT in the M61 galaxy (in the Virgo cluster) by  Koichi Itagaki \citep{nakano} and later spectroscopically classified as a type Iax SN on Oct. 31.20 UT 2014 by the Asiago Transient Classification Program \citep{orchner} using criteria defined in \citet{Foley13}. 
 The optical/near-infrared (NIR) and mid-infrared (MIR) properties of SN\,2014dt have been presented by \citet{Fox2015,Kawabata18},  while the optical properties of the host-galaxy environment have been studied by \citet{Lyman2018}. In addition to the progenitor limits of \citet{Foley2015a}, of particular interest is the finding of a very slow spectroscopic and photometric evolution of SN\,2014dt at $t\ge 100$ days post-maximum \citep{Kawabata18} and the presence of an MIR excess of emission \citep{Fox2015,Foley2016}. In analogy with the type Iax SN\,2005hk  \citep{Foley13}, this slow evolution can be interpreted as evidence for the presence of a persistent photospheric component  possibly formed within the wind launched by the surviving bound remnant WD \citep{Kawabata18}. The optical-NIR luminosity decline of SN\,2014dt is the slowest observed in well studied SNe Iax \citep{Kawabata18}.

In this paper we utilize deep X-ray and radio observations of the nearby type Iax SN\,2014dt in the first $\sim$50 days since explosion. X-ray and radio emission originate from particles accelerated at the SN shock to relativistic speeds (e.g., \citealt{Chevalier06}), as the blastwave plows through the circumstellar medium (CSM), which was sculpted by mass lost to the surroundings by the SN progenitor system. X-ray and radio observations of SNe from WD binary systems thus carry direct information about the mass-loss history of their stellar companions, and therefore their nature. This technique has been recently employed by \citet{Immler02,Margutti12, Chomiuk12, Horesh12,Russell12,Chomiuk2015}. No type Iax SN has ever been detected either in the radio or X-ray regime. From the most recent compilation by \citet{Chomiuk2015}, typical radio luminosity limits are $L_{\nu,R}\lesssim 1.81 \times  10^{25}\,\rm{erg\,s^{-1}Hz^{-1}}$ (at $\nu\sim5.9-8.4$ GHz). Typical X-ray limits are $L_x \lesssim 10^{39}\,\rm{erg\,s^{-1}}$ (this work). Our analysis of SN\,2014dt extends the range of investigated radio and X-ray luminosities to $L_{\nu,R}\sim (1.0 - 2.4) \times10^{25}\,\rm{erg\,s^{-1}Hz^{-1}}$ and $L_x\sim 10^{38}\,\rm{erg\,s^{-1}}$.

This paper is organized as follows. We present the X-ray and radio observations of SN\,2014dt in \S\ref{Sec:Obs}. We model the X-ray and radio observations in the context of IC and synchrotron emission, respectively, in \S\ref{Sec:IC} and \S\ref{Sec:RadioConstraints}. We discuss our constraints on the progenitor systems of type Iax SNe and thermonuclear explosions and present our conclusions and prospects for future X-ray and radio studies of type Iax SNe  in in \S\ref{Sec:disc}. Throughout the paper uncertainties are provided at the 68\% confidence level (c.l.) and upper limits at the $3\,\sigma$ c.l. (Gaussian equivalent) unless explicitly stated otherwise.

\section{Observations} 
\label{Sec:Obs}

The precise distance of M61 and the precise explosion date of  SN\,2014dt are unknown. \citet{Foley2015a} favor the distance of 12.3 Mpc  ("EPM method" \citet{bose2014distance}). However, \citet{Fox2015} adopt the distance of 19.3 Mpc ("Photospheric magnitude method" \citep{Rodriguez}).
 SN\,2014dt was discovered around peak optical brightness, so the explosion date is not well constrained. The time of maximum luminosity in the B-band was constrained to October 20, 2014 (MJD $\sim$ 56950) (\citealt{Fox2015,Foley2016,singh18,Kawabata18}). 
 In the following we conservatively derive and discuss our results for the entire range of distances between $d=$12.3-19.3 Mpc, and explosion epochs in the range MJD 56930-56940 (corresponding to a rise time between 10-20 days, in line with other type Iax SNe, e.g., \citealt{Magee16,Magee17,Jha2017}).

\subsection{X-ray Observations with the CXO}
\label{Sec:XrayObs}
\begin{figure}
\includegraphics[width=3.3 in]{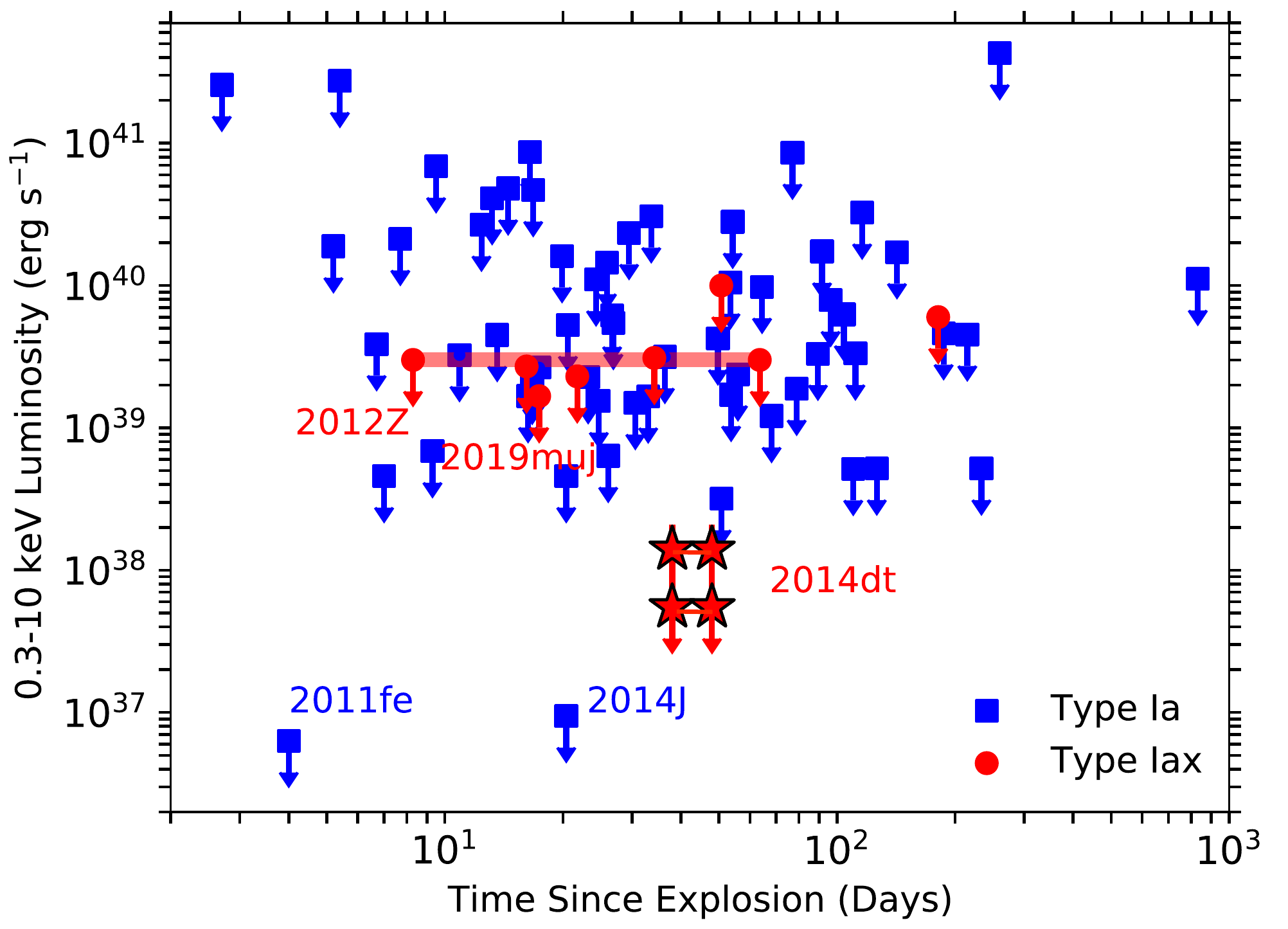}
\caption{CXO observations of SN\,2014dt (red stars) provide the deepest constraints on the X-ray luminosity of a type-Iax SN to date. The four red stars indicate the limits derived for SN\,2014dt taking the range of distance and explosion dates into account. We also plot the X-ray limits of type-Iax SNe 2012Z and 2019muj from \emph{Swift}-XRT observations.  Red (blue) points: X-ray limits from previous observations of type Iax (type Ia) SNe from \citet{Russell12}, \citet{Margutti12}, \citet{Margutti14c} and this work.}
\label{fig:xraydata} 
\end{figure}

We initiated deep X-ray observations of SN\,2014dt with the Chandra X-ray Observatory (CXO) on November 17th, 2014 under an approved DDT program 15508486 (PI Margutti), corresponding to 38-48 days since explosion. The data have been reduced with the CIAO software package (version 4.11), with calibration
database CALDB version 4.8.2. Standard ACIS data filtering has been applied. The total exposure time of our observations is 19.8\,ks. No X-ray source is detected at the SN position.
Assuming pure Poisson statistics we infer a 3$\sigma$ upper limit of $2.5 \times 10^{-4}\,\rm{c\,s^{-1}}$ in the 0.5-8 keV energy band, which implies an absorbed flux limit of $F_{x} < 3.5 \times 10^{-15} \rm{erg\, s^{-1} \, cm^{-2}}$ (0.3-10 keV) for a $F_{\nu} \propto \nu^{-1}$ non-thermal spectrum (which applies to both synchrotron and IC emission). The Galactic neutral hydrogen (NH) column density in the direction of SN\,2014dt is $1.7\times 10^{20} \rm{ cm^{-2}}$ \citep{Kalberla05}. The corresponding 0.3-10 keV luminosity limit for $d=12.3-19.3$ Mpc is  $L_x<(0.6-2.0)\times 10^{38}\,\rm{erg\,s^{-1}}$ (Fig. \ref{fig:xraydata}). 
The presence of diffuse soft X-ray emission from the host galaxy prevents us from reaching deeper limits.

\subsection{Radio Observations with the VLA} 
\label{Sec:RadioObs}

\begin{table*}
	\centering
	\caption{VLA observations of SN\,2014dt on November 7th 2014, corresponding to $\delta t $=28-38 days since explosion. For each observation we provide the range of luminosity density that corresponds to the range of distances 12.3-19.3 Mpc.}
	\label{tbl:values}
	\begin{tabular}{lccr}
		\hline
		\textbf{Central Freq.} & \textbf{Bandwidth} & \textbf{rms} & $\rm{L_{\nu}}$ \\
		{(GHz)} & {(GHz)} & {(mJy/beam)} & {$\rm{(erg\,s^{-1}\,Hz^{-1} )}$ }  \\ 
        \hline \hline
        4.8 & 2 & 0.033 & $<$(1.8 - 4.4)   $\times 10^{25}$ \\
        \hline
        7.5 & 2 & 0.018 & $<$(1.0 - 2.4)   $\times 10^{25}$ \\
        \hline
        20.0  & 4 & 0.019 & $<$(1.0 - 2.6) $\times 10^{25}$ \\
        \hline
        24.0  & 4 & 0.020 & $<$(1.1 - 2.7) $\times 10^{25}$ \\
		\hline
	\end{tabular}
\end{table*}

Deep radio observations were acquired with the Karl G. Jansky Very Large Array (VLA) with program VLA/14A-494 (PI A. Horesh) on November 7th 2014. The observations were taken at C, X, and K bands. These data were calibrated using the CASA pipeline version 1.3.1 \citep{casa}.   
We inspected the calibrated data and performed additional flagging in AIPS \citep{Greisen03}.  To minimize frequency-dependent artifacts in imaging, we split each of the C and K bands in two and imaged each half separately, producing four images centered at 4.8, 7.4, 20.0 and 24.0 GHz.  We used \texttt{difmap} for imaging \citep{Shepherd97}.  

\begin{figure}
\includegraphics[width=3.3in]{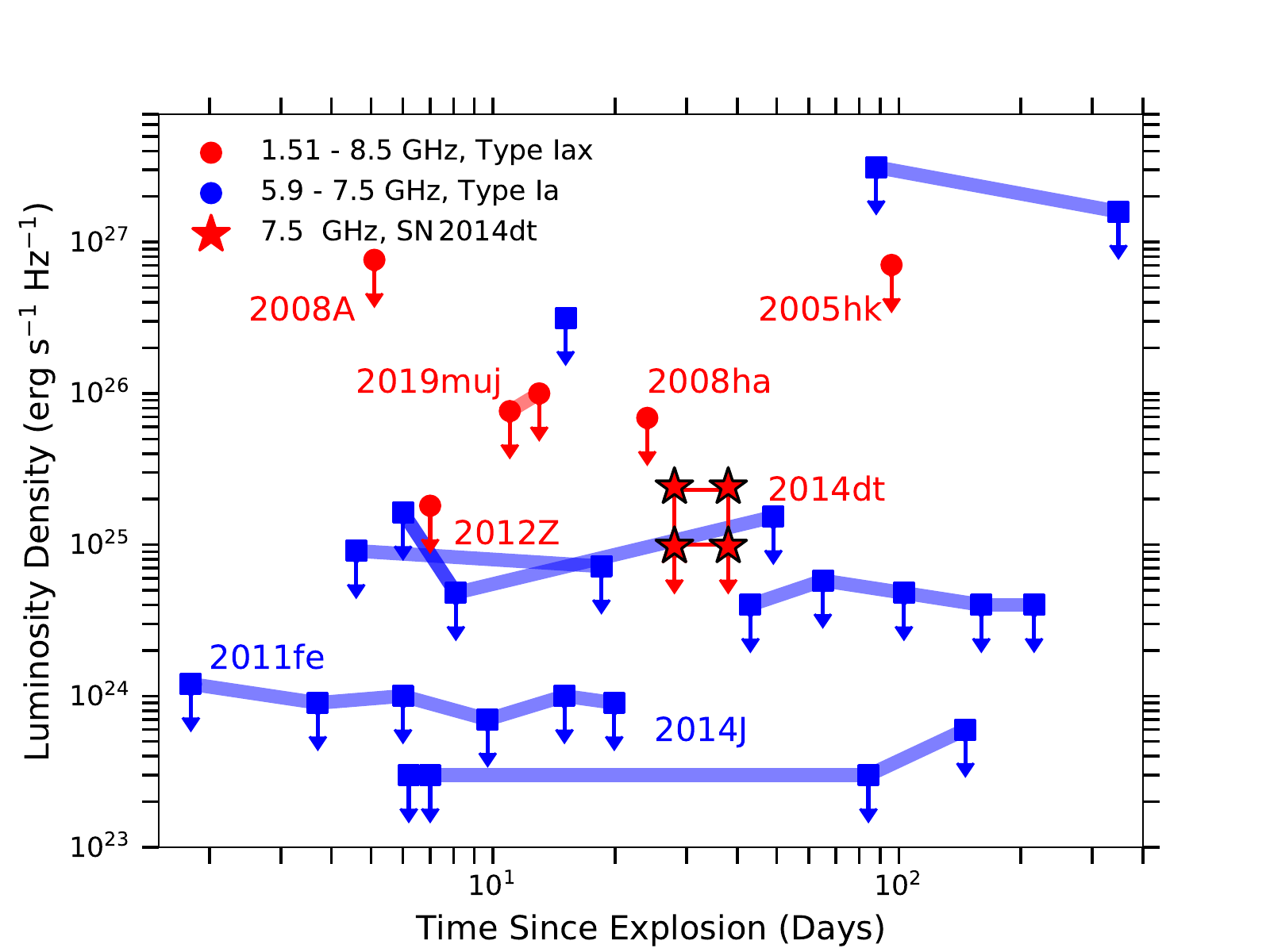}
\caption{Radio observations of SN\,2014dt (red stars) in the context of type Ia SNe (blue) and type Iax SNe (red) from \citet{Chomiuk2015}. Four red stars  limit  the  region  of  the  parameter  space  of  SN\,2014dt  depending  on the assumed distance and explosion epoch. Together with SN\,2012Z \citep{Chomiuk2015}, radio observations of SN\,2014dt provide the deepest constraints on radio emission from a type Iax SN at $t>10$ days. The radio limit on SN\,2019muj is from  \citet{2019ATel13105....1P}. } 
\label{fig:radiosample}
\end{figure}

Diffuse emission from the host galaxy dominates the images at 4.8 and 7.5 GHz.  To improve our sensitivity to point sources at these frequencies, we removed all baselines shorter than 0.01 M$\lambda$. The host galaxy flux does not contribute significantly to the 20 and 24 GHz images. All baselines were used to create images at these frequencies.  SN\,2014dt is not detected at any frequency.  We report the off-source root mean squared (rms) measurements for the images using natural weighting in Table \ref{tbl:VLA}. We compare the $3\sigma$ 7.5 GHz radio luminosity limits calculated for SN\,2014dt to a sample of type-Ia SNe and Iax SNe from \citet{Chomiuk2015} in Fig. \ref{fig:radiosample}. Together with SN\,2012Z, our radio limits on SN\,2014dt sample the lowest radio luminosities of type-Iax SNe.

\begin{figure*}
    \center{\includegraphics[width=7in]{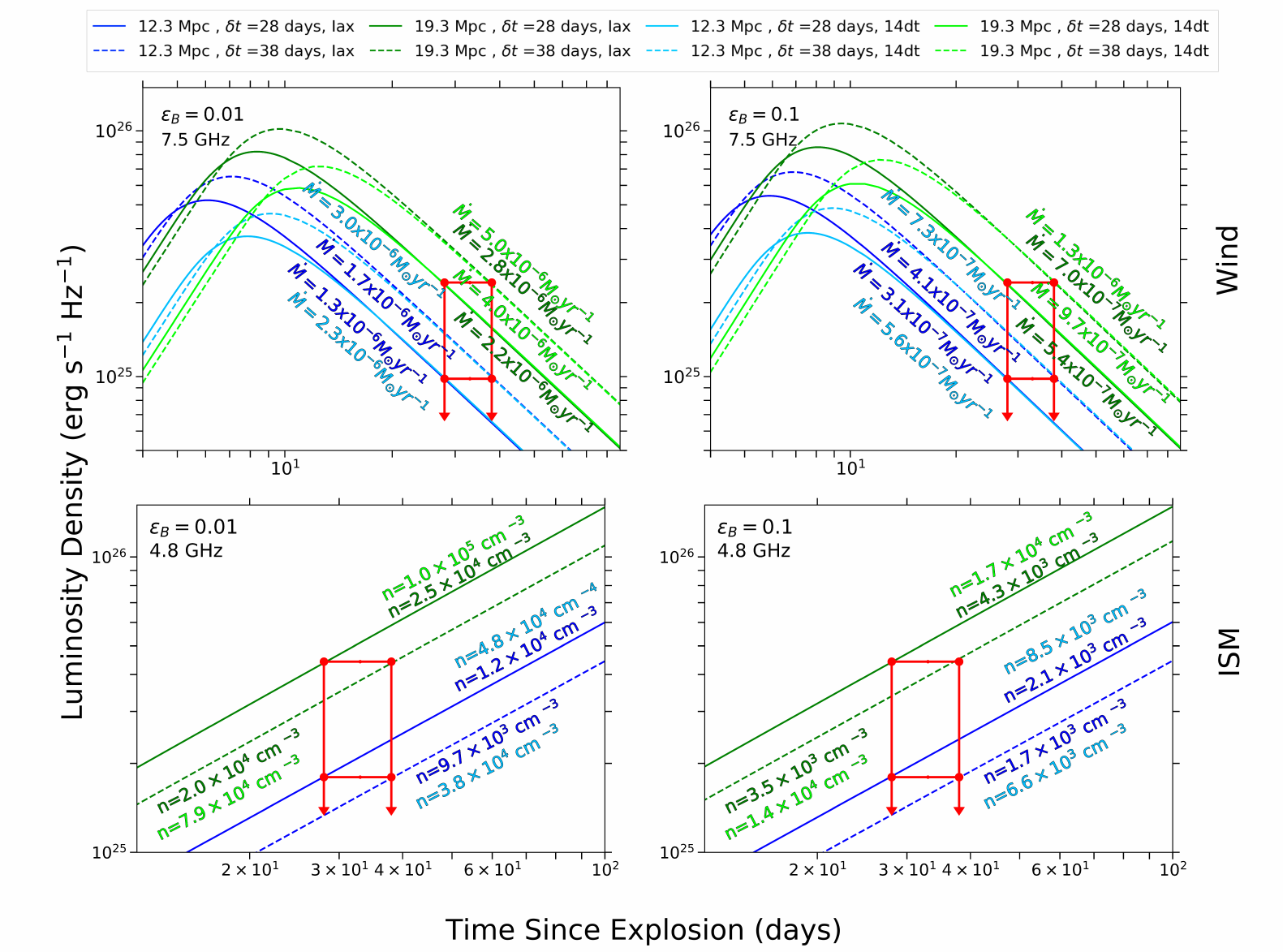}}
    \caption{Temporal evolution of the radio emission from a blast wave plowing through an ISM-like CSM for 4.8 GHz (lower panels) or wind-like CSM for 7.5 GHz (upper panels). Our calculations assume $\epsilon_e=0.1$ and $\epsilon_B=0.1$ (right panels) or $\epsilon_B=0.01$ (left panels), and adopt the wind velocity  $v_w=100\,\rm{km\,s^{-1}}$ to calculate $\dot M$. In dark shades of blue and green we show the radio luminosity for a ``typical'' type Iax SN explosion parameters where $M_{ej}=0.4\,\rm{M_{ch}}$ and $E_{k}=10^{50}\,\rm{erg}$. Lighter shades of blue and green represent the model for the lower $E_{k}$ and $M_{ej}$ values inferred for
    SN\,2014dt, which are  $M_{\rm{ej}}=(0.06-0.25)\,\rm{M_{ch}}$ and $E_{\rm{k}}=(0.07-0.42)\times 10^{50}\,\rm{erg}$ \citep{Kawabata18}. } 
    \label{Fig:radioconstraints}
\end{figure*}

\section{Constraints on the environment density from X-ray observations}
\label{Sec:IC}
X-ray emission generated by the upscattering of seed optical photospheric photons by a population of relativistic electrons accelerated at the SN shock (IC radiation) dominates  the early ($t\lesssim 40$ days) X-ray luminosity from SNe that explode in low-density media (like type Ia or type Ib/c SNe, e.g. \citealt{Chevalier06}). We independently verified this statement using the synchrotron modeling detailed in \S\ref{Sec:RadioConstraints} and found that the expected luminosity of synchrotron emission in the X-rays at these times is negligible compared to the the observed emission. We therefore continue under the assumption that at the time of our CXO observations of SN\,2014dt, IC dominates. After the SN optical peak, the IC emission became progressively fainter with time as the optical emission of the SN faded, and non-thermal synchrotron emission became the dominant X-ray component.

We adopt the IC formalism of \citet{Margutti12,Margutti14c}.
The IC emission depends on the explosion parameters (i.e. ejecta mass $M_{ej}$ and kinetic energy $E_k$), the outer density profile of the ejecta $\rho_{SN}\propto R^{-n}$, the environment density profile  $\rho_{CSM}\propto r^{-s}$, the energy spectrum of the accelerated electrons $N_e(\gamma)\propto \gamma^{-p}$, the fraction of postshock energy into relativistic electrons  $\epsilon_e$, and the SN bolometric luminosity $L_{bol}$. We assume $n\sim10$ as appropriate for compact stellar progenitors \citep{Chevalier06,Matzner99}, we employ a wind-density profile $s=2$ and we use $p=3$, as typically inferred from radio observations of SNe (e.g. \citealt{Soderberg06,Soderberg06b,Soderberg06c,Soderberg10}), where the IC luminosity is $L_{\rm{IC}}(t)\propto L_{bol}(t)$. We use the optical bolometric luminosity curve of SN\,2014dt derived by \citet{Kawabata18}.

\begin{figure*}
\includegraphics[width=7in]{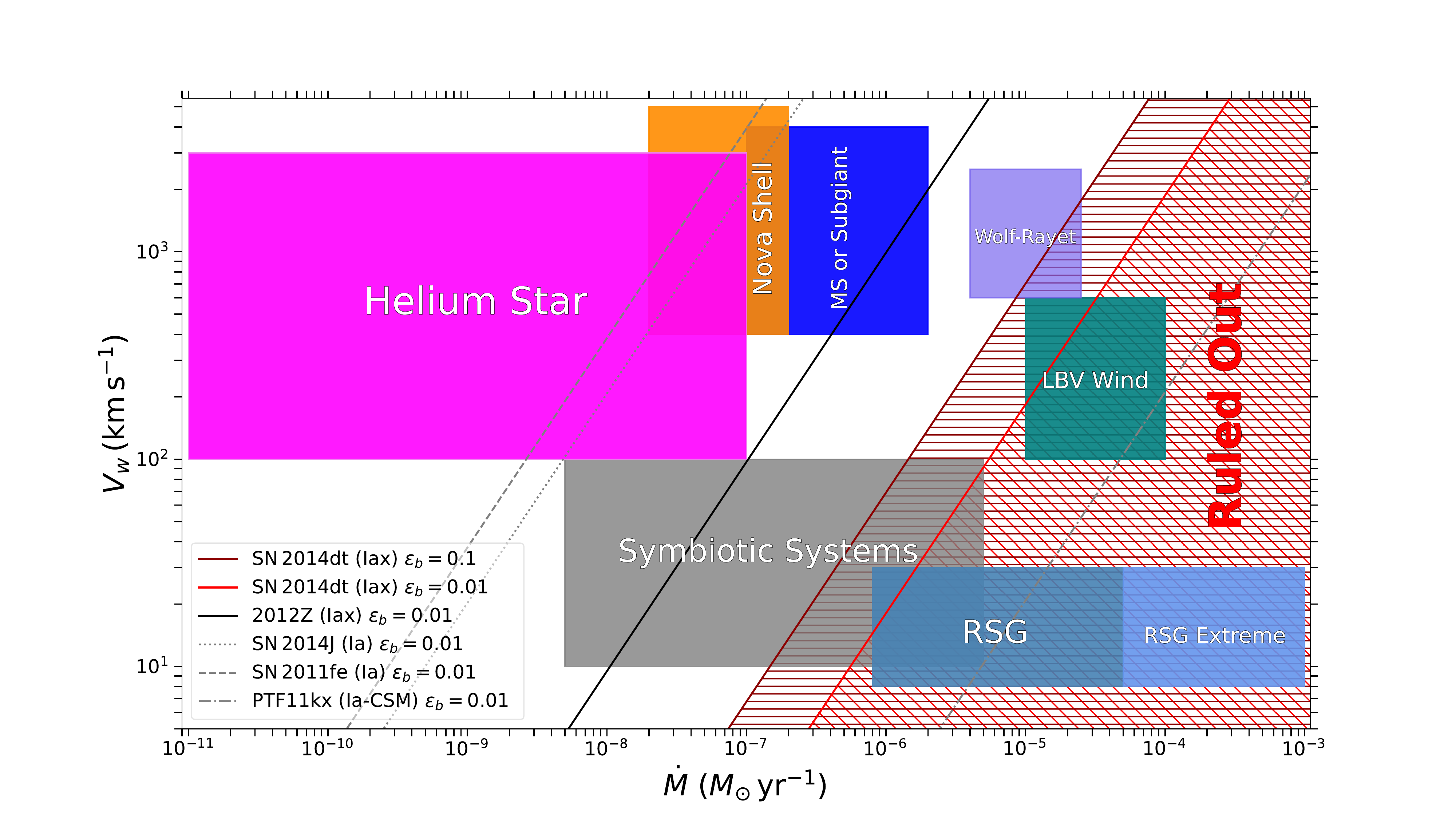}
\caption{Constraints on the parameter space of the mass-loss rate from the binary progenitor system $\dot M$ and velocity of the progenitor wind $v_w$. Hatched regions indicate the part of the parameter space that is ruled out. For SN\,2014dt we derive the most conservative $\dot{M}$ for $\epsilon_{B}$ = 0.01  and  $\epsilon_{B}$ = 0.1
as ${\dot{M}}<$  5.0 $\times 10^{-6} \rm{M_{\sun}\,yr^{-1}} (v_w/100\,\rm{km\,s^{-1}})$ and  ${\dot{M}}<$  1.3 $\times 10^{-6} \rm{M_{\sun}\,yr^{-1}}  (v_w/100\,\rm{km\,s^{-1}}) $, respectively.  
We compare these limits to those for $\rm{\epsilon_{B} = 0.01}$ of SNe Iax 2012Z \citep{Chomiuk2015} and 2019muj (derived in Appendix \ref{appendix:a} using the radio upper limits from \citealt{2019ATel13105....1P}), and SNe Ia 2014J \citep{Chomiuk14}, 2011fe \citep{Chomiuk12}, and the type Ia-CSM PTF11kx \citep{dilday2012ptf,Chomiuk2015}  where $\rm{\epsilon_{B} = 0.01}$. Using the observed mass-loss rates and wind velocities in massive stars from \citet{Chomiuk12} and \citet{terreran2019sn}, we find that a portion of the  parameter space associated with  giant star (GS) companions and Wolf-Rayet (WR) stars is ruled out for $\rm{\epsilon_{b} =0.1}$  for SN\,2014dt. For both values of $\epsilon_{B}$ we can rule out Red Super Giant Stars (RSG) and most of the Luminous Blue Variable (LBV) stars wind-like CSM parameter space. We do not constrain the parameter space of He-stars, Nova-like shells, or MS stars.}
\label{Fig:massloss}
\end{figure*}

We present the results of our IC modeling for two choices of the explosion parameters. First we assume the parameters inferred by \citet{Kawabata18} from the modeling of the post-peak optical/NIR light-curve of SN\,2014dt. \citet{Kawabata18} derive $M_{\rm{ej}}=0.06-0.25\,\rm{M_{ch}}$ (with $\rm{M_{ch}}=1.4\,\rm{M_{\sun}}$) and $E_{\rm{k}}=(0.07-0.42)\times 10^{50}\,\rm{erg}$.
The low ejecta mass and kinetic energy are consistent with a surviving remnant scenario. For these parameters we find $\dot M<3.5\times 10^{-1}\,\rm{M_{\sun}yr^{-1}}$ for $v_w=100\,\rm{km\,s^{-1}}$. Second, we provide a calculation of the mass-loss limit assuming  the typical explosion parameters of type Iax SNe as a class, i.e.  $M_{ej}\sim 0.4\,\rm{M_{ch}}$  and $E_{k}\sim10^{50}\,\rm{erg}$. For these parameters we infer $\dot M<1.1\times 10^{-1}\,\rm{M_{\sun}yr^{-1}}$ for $v_w=100\,\rm{km\,s^{-1}}$.  In the first scenario the inferred mass-loss limits are slightly less constraining as there is less ejecta mass carrying energy to power the emission (Fig. \ref{Fig:massloss}).  

\section{Constraints on the environment density from Radio observations}
\label{Sec:RadioConstraints}
In stellar explosions the thermal optical emission traces the slow moving ejecta ($v\leq 10000\,\rm{km\,s^{-1}}$) to which the bulk of the kinetic energy is coupled, while radio observations probe the fastest ejecta ($v\geq0.1c$). The radio emission originates from the interaction of the fastest SN ejecta with the local environment that results from the recent mass loss from the progenitor star. 

The observed synchrotron radiation depends on (i) the explosion parameters ($M_{ej}$, $E_k$), (ii) on the density of the surrounding medium ($\rho_{CSM}\propto r^{-s}$ with $s=2$ for a wind profile and $s=0$ for an interstellar medium ISM-like CSM profile), (iii) on the fraction of post-shock energy into magnetic fields and  relativistic electrons ($\epsilon_B$ and $\epsilon_e$, respectively), and (iv) on the electron distribution  $N_e(\gamma)\propto \gamma^{-p}$. We adopt the formalism of \citet{Chomiuk2015} for $p=3$ and infer a density of the environment assuming both ISM and wind-like environments. Our results are reported in Table \ref{tbl:VLA}.  As before we
estimate the density and mass-loss rate limits both for the explosion parameters of SN\,2014dt and for typical SN Iax explosion parameters to allow a direct comparison to the results presented by \citet{Chomiuk2015}.

\begin{table*}

\caption{Limits on the environment density (for an ISM-like CSM) and  progenitor mass-loss rate of the pre-explosion (for a wind-like CSM) derived from the radio observations of Table \ref{tbl:values} and the modeling of \S\ \ref{Sec:RadioConstraints}. We present our inferences for the environment properties of SN\,2014dt using explosion parameters derived based on the modeling of the optical bolometric light-curve of SN\,2014dt by \citet{Kawabata18} ($M_{\rm{ej}}=$(0.06-0.25) $\rm{M_{ch}}$, $E_{\rm{k}}=$(0.07-0.42)$\times 10^{50} \rm{erg}$) as well as explosion parameters  that are typical of Iax SNe ($M_{\rm{ej}}= 0.4\ \rm{M_{ch}}$, $E_{\rm{k}}=10^{50} \rm{erg}$ ) or  to allow a direct comparison to the limits presented in \citet{Chomiuk2015}. For the mass-loss rate limits we adopt $v_{w}= $ 100 $ \rm{km\,s^{-1}}$.}

\centering 
\begin{tabular}{c|cc|cc}
\hline 
\textbf{CSM type } & \multicolumn{2}{|c|}{\textbf{Typical SN Iax}}& 
\multicolumn{2}{c}{\textbf{SN\,2014dt}}  \\
\cline{2-5}
& $\epsilon_B=0.1$ & $\epsilon_B=0.01$ & $\epsilon_B=0.1$ & $\epsilon_B=0.01$ \\
\hline
$n$ ($\rm{cm^{-3}}) <$ & $4.3 \times 10^{3}$ & $2.5 \times 10^{4}$ & $1.7 \times 10^{4}$ & $1.0 \times 10^{5}$ \\
$\dot M $ $(\rm{M_{\sun}yr^{-1}}) <$ & $7.0 \times 10^{-7}$ & $2.8 \times 10^{-6}$ & $1.3 \times 10^{-6}$ & $5.0 \times 10^{-6}$ \\ 
\hline

\end{tabular}
\label{tbl:VLA}
\end{table*}

\begin{figure*}
    \center{\includegraphics[scale=.69]{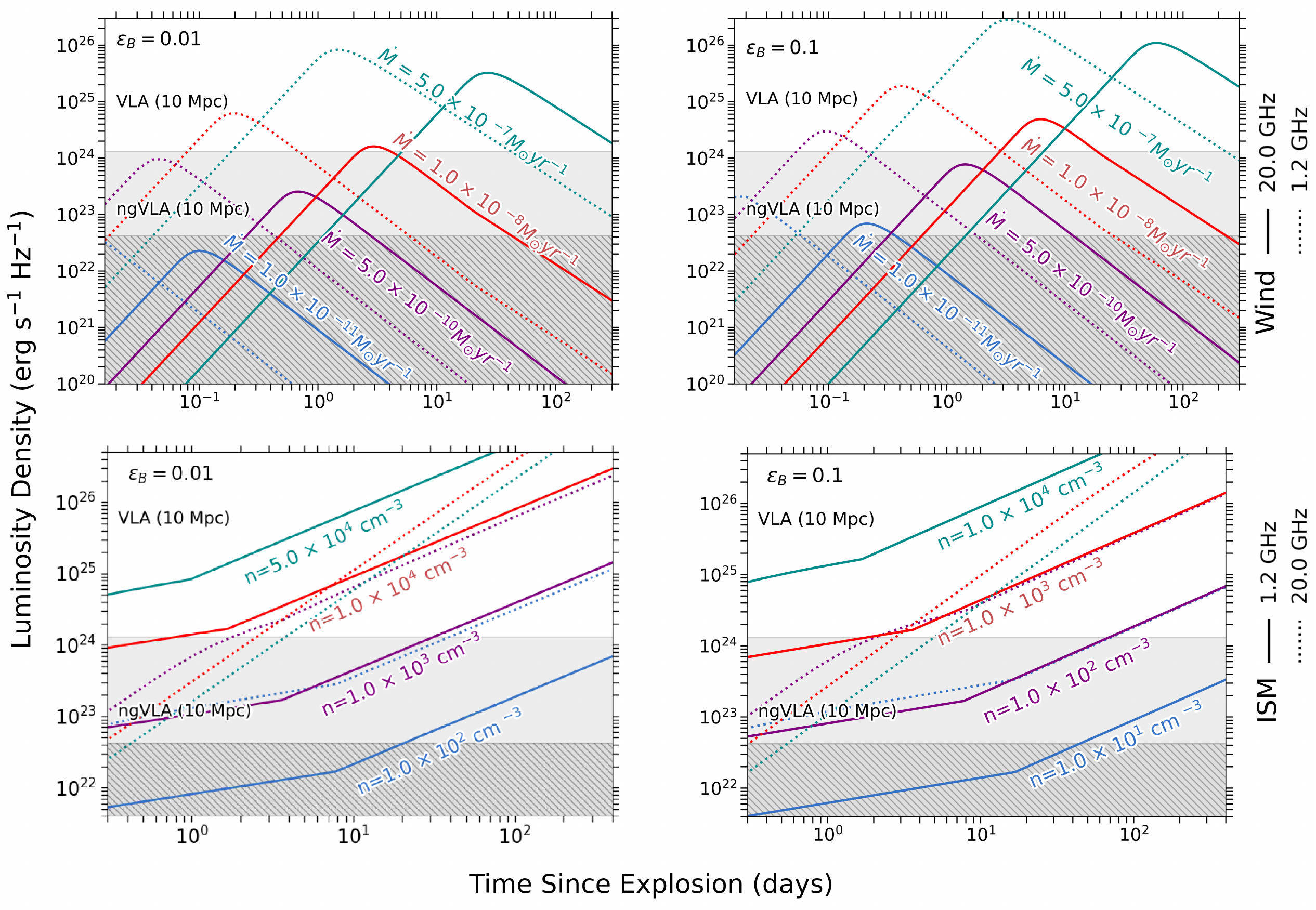}}
    \caption{Radio light-curves at 1.2 GHz (thick lines) and 20 GHz (dotted lines) of type-Iax SNe with explosion parameters similar to SN\,2014dt and a wind-like (upper panels) or ISM-like (lower panels) CSM. Different colors are used for different mass-loss values (for the wind scenario) or density (for the ISM scenario). We adopt $v_w=100\,\rm{km\,s^{-1}}$. The right panels adopt an optimistic $\epsilon_B=0.1$, while the left panels use the fiducial value of $\epsilon_B=0.01$ value. The black region indicates the region which can not be probed by either the VLA or ngVLA, while the light grey and white region show the typical luminosity limit (3$\sigma$) for
    $\sim$ 1 hr of ngVLA and VLA time at $\lesssim$ 20 GHz \citep{carilli2015next} respectively for a SN at 10 Mpc. This exercise shows that even in the case of a very nearby SN, the improved sensitivity of the ngVLA is necessary to probe the very low mass-loss rates of He stars that might be the companions of WDs progenitors of Iax SNe.}
    \label{Fig:vla} 
\end{figure*}

\section{Discussion and conclusions}
\label{Sec:disc}

We have presented a sensitive search for X-ray  and radio emission (Figs. \ref{fig:xraydata}-\ref{fig:radiosample}) from the very nearby type Iax SN\,2014dt. Because of the presence of diffuse X-ray emission from the host galaxy, we found that the X-ray limits were not as constraining as the radio limits, and we therefore focus our discussion on the results from the radio modeling of \S\ \ref{Sec:RadioObs}. 

Radio observations of SN\,2014dt allowed us to place tight constraints on the environment density of a type Iax SN. For the progenitor system of SN\,2014dt we find that the most conservative $\dot{M}$ limits for $\epsilon_{B}$ = 0.01 and  $\epsilon_{B}$ = 0.1 are ${\dot{M}} < 5.0 \times 10^{-6} M_{\sun}\,yr^{-1} (v_w/100\,\rm{km\,s^{-1}}) $ and  ${\dot{M}}< 1.3 \times 10^{-6} M_{\sun}\,yr^{-1} (v_w/100\,\rm{km\,s^{-1}}) $, respectively. For standard assumptions on the SN ejecta profile and shock dynamics of \citet{Chomiuk2015}, the forward shock radius for a wind-stratified CSM environment is:

\begin{equation}
R_{FS} = (5.0 \times 10^{15} {\rm cm})\,M_{\rm ej,~Ch}^{-0.32}\,E_{k,~51}^{0.44}\,A_{*}^{-0.12}\,t_{10}^{0.88},
\label{equation:rfs}
\end{equation}

where $t_{10}$  is the time since explosion normalized to 10 days, $A_{*} = (A/5 \times 10^{11}) \rm{g \, cm^{-1}}$ where $A_{*} = 1$ corresponds to $\rm{\dot{M}/v_{w} = 10^{-6} \frac{M_{\sun}yr^{-1}}{100\,km\,s^{-1}}}$ \citep{Chevalier06}, $\rm{M_{ej,Ch}}$ is the ejecta mass in units of Chandrasekhar mass and $E_{k,51}$ is the kinetic energy in units of $10^{51}$ erg. We assume solar abundance for the CSM with mean molecular weight $\mu $= 1.4 \citep{Chomiuk2015}. Based on Equation \ref{equation:rfs} the mass-loss limits above imply forward shock radii $R_{FS} \gtrsim (5-9)\times 10^{15}\,\rm{cm}$ at 28-38 days since explosion.

We put our inferences into the context of the mass-loss parameter space of potential stellar companions to WDs in a single degenerate progenitor system  in Fig. \ref{Fig:massloss}. Our results exclude some symbiotic systems, RSG, LBV winds, and WR stars (Fig. \ref{Fig:massloss}). This result is consistent with the constraints placed on the progenitor of SN\,2014dt using Hubble Space Telescope (HST) pre-explosion images \citep{Foley2015a}.  Specifically, WR stars are disfavored by HST observations as WR stars typically reside in regions with high density of bright stars, whereas SN\,2014dt exploded in a sparsely populated stellar region ($\sim$100 pc from any detected sources, and more than $\sim$200 pc from any particularly bright source, \citealt{Foley2015a}).

SN\,2014dt showed evidence for an excess of emission at NIR-MIR wavelengths \citep{Fox2015}. Understanding the physical origin of this excess might offer additional clues about the progenitor system of SN\,2014dt. At $\sim$100 days after  maximum optical light, \citet{Fox2015} found that the emission from SN\,2014dt  begins to exhibit redder colors than the type Iax SNe 2005hk and 2012Z, reaching a color $(B-V)\sim 1.5$ mag at $\sim300$ days. A MIR excess of emission  was also detected about a year after explosion, which \citet{Fox2015} attributed to $\sim 10\rm{^{-5} M_{\sun}}$ of either optically thin newly formed dust \textit{or} optically thin pre-existing circumstellar dust, which was estimated to be at a radius $ \rm{r_{bb} \ge 3 \times 10^{15}\,cm} $. The latter scenario would support a single degenerate progenitor system with an inferred mass-loss rate of $\rm{< 10^{-6} M_{\sun}yr^{-1}}$ for a wind velocity  v$_{w}$ = 100 $\rm{km\,s^{-1}}$ \citep{Fox2015}.  Our radio observations rule out the presence of the dust shell at radii smaller than $\sim 10^{16}\,\rm{cm}$ for $\epsilon_B\ge 0.1$ but do not allow us to probe the entire parameter space associated with the dust-shell scenario. 

\citet{Foley2016} found the possibility of \textit{pre-existing} dust to be unlikely due to  low reddening of SN\,2014dt at early times. In addition, strong limits on narrow  absorption lines in the spectra of SN\,2014dt indicated a gas-poor environment \citep{Foley2015a} and spectra of SN\,2014dt  indicated no circumstellar interaction \citep{Foley2016}. 
\citet{Foley2016} also found no observational evidence for temporal evolution of the line profiles (i.e. extinction of the redshifted  emission line profile, which is a telltale feature of dust formation, e.g. \citealt{smith2008dust}). There were also no clear changes of the forbidden-line shapes from +270 days and +410 days, as well as no evidence for additional reddening which makes the newly-formed dust scenario unlikely. An alternative explanation of the MIR excess of emission  that is consistent with the observational evidence above is that of an optically thick super-Eddington wind launched by a bound remnant that survived the explosion. This scenario is supported by observations of a long-lasting photosphere and low photospheric velocities (\citealt{Foley2016}; see however \citealt{Kawabata18}). Our radio observations cannot probe the presence of this dense wind, since this wind was launched after explosion.

We conclude with one consideration on the future of radio studies of subluminous thermonuclear explosions. Our analysis of SN\,2014dt has shown how even in the case of the most nearby type Iax SNe, deep VLA observations acquired at relatively early times are unable to probe the parameter space of He-star progenitors. This conclusion applies to another recently-discovered nearby type-Iax SN\,2019muj, as we quantitatively demonstrate in Appendix \ref{appendix:a}. The significantly improved sensitivity of the next generation Very Large Array (ngVLA, \citealt{carilli2015next}), will help explore the parameter space of the very low-density media that might be typical of thermonuclear explosions down to unprecedented limits (Figure \ref{Fig:vla}).

\bigskip
\bigskip\bigskip\bigskip
%Margutti's group
Margutti's team is partially supported by the Heising-Simons Foundation under grant \#2018-0911 (PI: Margutti). R.M. is grateful to KITP for hospitality during the completion of this paper. This research was supported in part by the National Science Foundation under Grant No. NSF PHY-1748958. R.M.~acknowledges support by the National Science Foundation under Award No. AST-1909796 and AST-1944985. Raffaella Margutti is a CIFAR Azrieli Global Scholar in the Gravity \& the Extreme Universe Program, 2019, and a A.~P. Sloan fellow in Physics 2019.  C.S. is grateful to the Integrated Data-Driven Discovery in Earth and Astrophysical Sciences (IDEAS) of the National Science Foundation (NSF). W.J-G is supported by the National Science Foundation Graduate Research Fellowship Program under Grant No.~DGE-1842165 and the IDEAS Fellowship Program at Northwestern University. W.J-G acknowledges support through NASA grants in support of {\it Hubble Space Telescope} program GO-16075. A.H. is grateful for the support by grants from the Israel Science Foundation, the US-Israel Binational Science Foundation (BSF), and the I-CORE Program of the Planning and Budgeting Committee and the Israel Science Foundation. 

The UCSC transient team is supported in part by National Science Foundation (NSF) grant AST-1518052, the Gordon \& Betty Moore Foundation, the Heising-Simons Foundation, and by a fellowship from the David and Lucile Packard Foundation to R.J.F. 

%Swift XRT
 This research has made use of the XRT Data Analysis Software (XRTDAS) developed under the responsibility of the ASI Science Data Center (ASDC), Italy.
 %VLA-NRAO 
 The National Radio Astronomy Observatory is a facility of the National Science Foundation operated under cooperative agreement by Associated Universities, Inc.  Fundamental Research.
 %Chandra
 We thank the entire Chandra team for carrying out our DDT observations, and the Chandra X-ray Center, which is operated by the Smithsonian Astrophysical Observatory for and on behalf of the National Aeronautics Space Administration under contract NAS8-03060.


\bibliographystyle{mnras}
\bibliography{main.bib} % if your bibtex file is called example.bib

%%%%%%%%%%%%%%%%% APPENDICES %%%%%%%%%%%%%%%%%%%%%

\appendix

\newpage 
\section{Sensitivity of current (VLA) and future (SKA and ngVLA) radio telescopes} 

\begin{table*}
	\centering
	\caption{Sensitivity of current (VLA) and future (SKA and ngVLA) radio telescopes at representative frequencies that are relevant to Fig. \ref{Fig:vla}. The rms sensitivity is provided for a 1-hr on source observation and the luminosity limits have been calculated as 3$\times$RMS at the nominal distance of 10 Mpc.}
	\label{tab:vla}
	\begin{tabular}{lccr}
		\hline
		{\textbf{Telescope}} & { \textbf{Central Freq.}} & {\textbf{rms}} & {\textbf{Lum. Density Limit}} \\  
        & {(GHz)} & {($\mu$Jy/beam)} & {$\rm{(erg\,s^{-1}\,Hz^{-1} )}$ } \\
		\hline \hline

		\textbf{ngVLA}$^a$ & 2.4  & 0.38   &  1.4 $\times 10^{23}$\\
                       & 16.4 & 0.20   &  7.2 $\times 10^{22}$ \\
        \hline 
        \textbf{SKA}$^b$   & 1.4  &  2.0   & 7.2 $\times 10^{23}$\\
                       & 12.5 &  1.2   & 4.3 $\times 10^{23}$\\ 
        \hline
        \textbf{VLA}   & 1.4  & 8.0    &  2.9 $\times 10^{24}$\\                     
                       & 22.0 & 3.7    &  1.3 $\times 10^{24}$  \\     
		\hline
		\multicolumn{4}{l}{\footnotesize{$^a$https://ngvla.nrao.edu/page/performance}}\\
		\multicolumn{4}{l}{\footnotesize{$^b$https://www.skatelescope.org}}\\
	\end{tabular}
\end{table*}
\label{appendix:a}

\section{Inferences on the progenitor mass-loss history of the type-Iax SN2019muj}
SN\,2019muj was discovered on 7 August 2019, UT 09:36 (MJD 58702.4) by \citet{brimacombe2019asassn}. It was classified as a type-Iax SN at the distance of 25.8 Mpc \citep{2019ATel13105....1P}.  
\citet{2019ATel13105....1P} found no evidence of radio emission from SN\,2019muj 
and inferred  a  3$\sigma$ upper limit of {$F_{\nu}<$} 96$\rm{\mu}\,$Jy/beam and {$F_{\nu}<$} 126 $\rm{\mu\,}$Jy/beam at 5.08 and 1.51 GHz, respectively. The corresponding upper limits on the monochromatic luminosity are $L_{\nu}<$ 7.6 $\times 10^{25}$ erg/s/Hz and $L_{\nu}<1.0 \times 10^{26}$ erg/s/Hz at 5.08 and 1.51 GHz, respectively, for an assumed distance of 25.8 Mpc. Using these measurements \citet{2019ATel13105....1P}  reported upper limits on the mass-loss rate of the supernova progenitor of $\dot M<$ 6.1 $\times 10^{-8}$ $\rm{M_{\sun} yr^{-1}}$ and $\dot M<$ 5.6 $\times 10^{-8}$ $\rm{M_{\sun} yr^{-1}}$  (3-$\sigma$ c.l.) using observations at 5.08 and 1.51 GHz, respectively, for an assumed wind speed of $v_w=$100 km/s and optically thin synchrotron emission.

\begin{figure*}
    \center{\includegraphics[scale=.7]{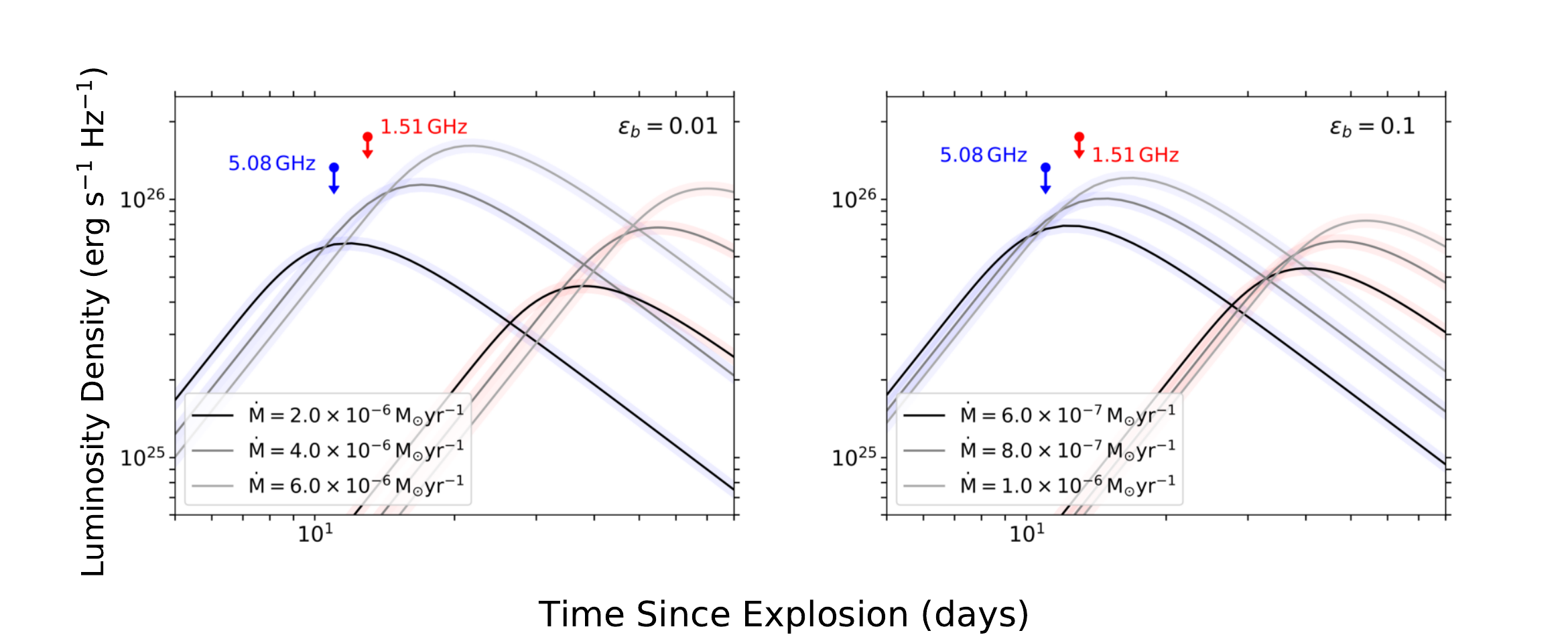}}
    \caption{Temporal evolution of the specific radio luminosity powered by synchrotron emission at 1.5 GHz (brown lines) and 5.1 GHz (blue lines) for three representative $\dot M$ values, $\epsilon_{B}$= 0.01 (\emph{left panel}) and  $\epsilon_{B}$ = 0.1 (\emph{right panel}) for SN\,2019muj. We use the distance and explosion parameters inferred by \citet{barna2020sn}  ($d = 34$ Mpc, ejecta mass $M_{ej}$ = 0.12 $M_{ch}$ and kinetic energy $E_{k}$ = 0.04 $\times 10^{51}$ erg), and adopt a wind velocity $\rm{v_{w}=100\,km\,s^{-1}}$. Red and blue arrows: upper limit on the radio emission from SN\,2019muj presented by  \citet{2019ATel13105....1P}. The radio upper-limits do not rule out any of the synchrotron models (i.e. there is no model that violates the limits, irrespective from the mass-loss rate value is), and hence do not constrain the environment properties of SN\,2019muj.}
    \label{Fig:19mujwind}
\end{figure*}

\begin{figure*}
    \center{\includegraphics[scale=.68]{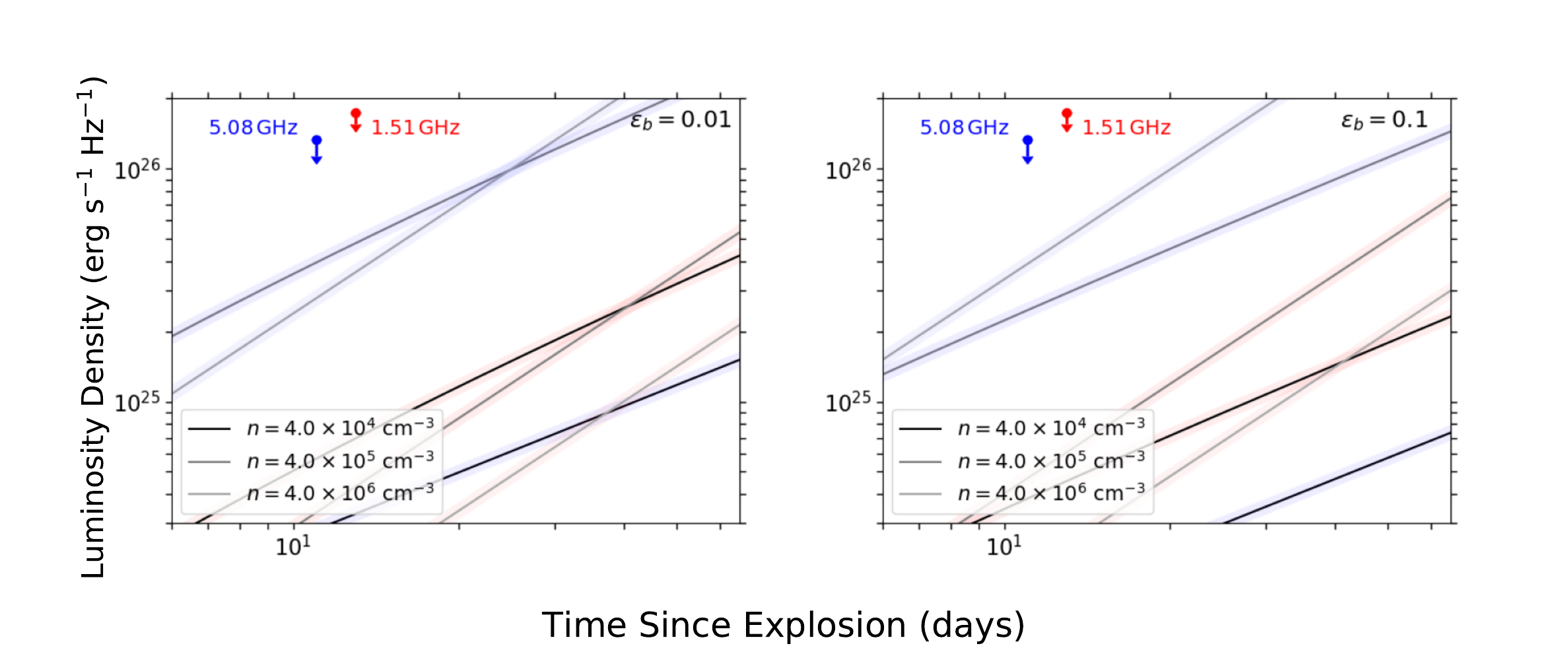}}
    \caption{Same as Fig. \ref{Fig:19mujwind} for an ISM-like CSM around SN\,2019muj. Similarly to the wind-scenario, the radio upper-limits do not rule out any of the synchrotron models, and hence do not constrain the environment properties of SN\,2019muj.} 
    \label{Fig:19mujISM} 
\end{figure*}

Here we adopt the revised distance to SN\,2019muj calculated by \citet{barna2020sn} based on the association with the host galaxy  VV\,525 for which a distance was calculated with the Tully-Fisher method  $d=34.1\,\rm{Mpc}$. We further adopt the explosion parameters $\rm{M_{ej}}= 0.12 \,\rm{M_{ch}}$, kinetic energy  $E_{k}= 0.04  \times 10^{51}\,\rm{erg}$ from the modeling of the  optical light-curve of SN\,2019muj  presented by \citet{barna2020sn}.  Applying the synchrotron formalism of \S\ref{Sec:RadioConstraints} we do not confirm the inferences from \citet{2019ATel13105....1P}, but rather conclude that their radio limits of SN\,2019muj presented by \citet{2019ATel13105....1P} do not constrain the mass-loss parameter space in either the ISM or the wind-like scenarios as we show in Figs. \ref{Fig:19mujwind}-\ref{Fig:19mujISM}. Specifically, while larger density ($n$) values for the ISM-like CSM or, equivalently, larger $\dot M$ values of the wind-like CSM model would be more luminous, denser media also cause the radio emission to peak at later times.
Deep radio observations of SN\,2019muj acquired at later times can potentially constrain the mass-loss parameter space of this explosion.

\label{appendix:b}

%%%%%%%%%%%%%%%%%%%%%%%%%%%%%%%%%%%%%%%%%%%%%%%%%%

% Don't change these lines
\bsp	% typesetting comment
\label{lastpage}
\end{document}